\documentclass[12pt,a4paper,fleqn]{article}

\usepackage[textheight=23cm,textwidth=15cm]{geometry}
\usepackage{amsmath}
\usepackage{graphicx}
\usepackage[american]{babel}
\usepackage{here}
\usepackage[articletitle=true]{achemso}

\begin{document}

\begin{center}

{\LARGE\bf
Expansive Quantum Mechanical Exploration of Chemical Reaction Paths
}

\vspace{1cm}

{\large
Alberto Baiardi,
Stephanie A. Grimmel,
Miguel Steiner,
Paul L. T\"urtscher,
Jan P. Unsleber,
Thomas Weymuth, and
Markus Reiher
}\\[4ex]

Laboratory of Physical Chemistry, ETH Zurich, \\
Vladimir-Prelog-Weg 2, 8093 Zurich, Switzerland \\
e-mail: markus.reiher@phys.chem.ethz.ch

November 12, 2021

\vspace{.43cm}
\textbf{Conspectus}
\end{center}
\vspace*{-.41cm}
{\small
Quantum mechanical methods have been well established for the elucidation 
of reaction paths of chemical processes and for the explicit dynamics of molecular systems.
While they are usually deployed in routine manual calculations on reactions
for which some insights are already available (typically from experiment), new algorithms and continuously
increasing capabilities of modern computer hardware allow for exploratory
open-ended computational campaigns that are unbiased and therefore enable unexpected
discoveries. Highly efficient and even automated procedures facilitate systematic approaches toward the
exploration of uncharted territory in molecular transformations and dynamics. 
In this work, we elaborate on such explorative approaches that range from reaction network explorations with
(stationary) quantum chemical methods to explorative molecular dynamics and migrant wave-packet dynamics.
The focus is on recent developments that cover the following strategies.
(i) Pruning search options for elementary reaction steps by heuristic rules based on the first principles of quantum mechanics:
Rules are required for reducing the combinatorial explosion of potentially reactive atom pairings
and rooting them in concepts derived from the electronic wave function makes them applicable to any molecular system.
(ii) Enforcing reactive events by external biases:
Inducing a reaction requires constraints that steer and direct elementary-step searches, which can be formulated in terms of 
forces, velocities, or supplementary potentials.
(iii) Manual steering facilitated by interactive quantum mechanics:
As ultrafast quantum chemical methods allow for real-time manual interactions with molecular systems,
human-intuition guided paths can be easily explored with suitable human-machine interfaces.
(iv) New approaches for transition-state optimization with continuous curve representations can provide stable 
schemes to be driven in an automated way by allowing
for an efficient tuning of the curve's parameters (instead of a manipulation of a collection of structures along the path),
and (v) reactive molecular dynamics and direct wave-packet propagation exploit the equations of motion
of an underlying mechanical theory (usually, classical Newtonian mechanics or Schr\"odinger quantum mechanics).
Explorative approaches are likely to replace the current state of the art in computational
chemistry because they reduce the human effort to be invested in reaction path elucidations,
they are less prone to errors and bias free, and they cover more extensive regions of the relevant configuration 
space. As a result, computational investigations that rely on these techniques are more likely to deliver surprising discoveries.
}


\section*{Key References}
\begin{itemize}
\item Haag, M.~P.;  Reiher, M. Real-Time Quantum Chemistry. \textit{Int.~J.~Quantum Chem.}~\textbf{2013}, \textit{113}, 8--20.\cite{Haag2013}.
In this work, we proposed the concept of real-time quantum chemistry which considers the ultra-fast calculation of energies and forces as a basis for interactive and automated quantum mechanics realized by our research group in the past decade.
The concept of interactive chemical reactivity exploration takes advantage of human chemical intuition to explore molecular systems.
\item Bergeler, M.; Simm, G.~N.; Proppe, J.; Reiher, M. Heuristics-Guided Exploration of Reaction Mechanisms. \textit{J.~Chem.~Theory Comput.}~\textbf{2015}, \textit{11}, 5712--5722.\cite{Bergeler2015}
This work advocates for automated exploration algorithms of chemical reaction networks that are agnostic with respect to the composition of the molecular systems. Therefore, they aim at taming the combinatorial explosion of options by what we call first-principles heuristics where the electronic wave function is interpreted to identify reactive, but more importantly unreactive sites.
\item Vaucher, A.~C.; Reiher, M. Minimum Energy Paths and Transition States by Curve Optimization. \textit{J.~Chem.~Theory Comput.} \textbf{2018}, \textit{14}, 3091--3099.\cite{Vaucher2018}
This work is part of our efforts to devise stable algorithms for automated and interactive reaction mechanism exploration. It introduces an optimization of reaction paths represented as continuous curves rather than as a series of distinct structures, making path optimizations fast and efficient.
\item Baiardi, A.; Reiher, M. Large-Scale Quantum Dynamics with Matrix Product States. \textit{J.~Chem.~Theory Comput.}~\textbf{2019}, \textit{15}, 3481--3498.\cite{Baiardi2019}
This article presents our implementation of the time-dependent density matrix renormalization algorithm, which allowed us to study molecular systems that are challenging for traditional state-of-the-art quantum dynamics algorithms.
\end{itemize}

\section{Introduction}
A key question in chemistry is to what products some given reactants may react and which of these are the most likely ones.
Driving forces for these processes can be thermal motion, excitation by light, or even directed processes such as mechanochemical
ones.
From the point of view of potential energy surface (PES) exploration, this requires us to examine what other local minima 
representing reaction products and stable intermediates can be reached from a given minimum (representing the reactants)
and how high the barriers are that are associated with these transformations \cite{Bofill2020}.
Exhaustive calculations of an entire PES are, however, unfeasible for all but the tiniest molecules.
For this reason, computational analyses are typically limited to the inspection of
selected (collective) transformation coordinates defining specific reaction coordinates along a PES.

Owing to the advances in ubiquitous computing and comparatively easy accessibility of reliable quantum chemical methods,
the state of the art of reaction mechanism elucidation has been pushed toward the massive exploration of vast reaction
networks in the past decade \cite{Vazquez2018,Dewyer2018,Simm2019,Unsleber2020,Maeda2021}. Whereas these 
developments mostly focused on time-independent quantum chemical methods, the exploratory nature of computational approaches
has much earlier been exploited in molecular dynamics simulations.
Here, the necessity to enhance rare event
sampling posed a severe problem that led to the development of a broad range of algorithms tailored
to produce a rare reactive event in a simulation that is feasible.

\begin{figure}[H]
    \centering
    \includegraphics[width=0.99\textwidth]{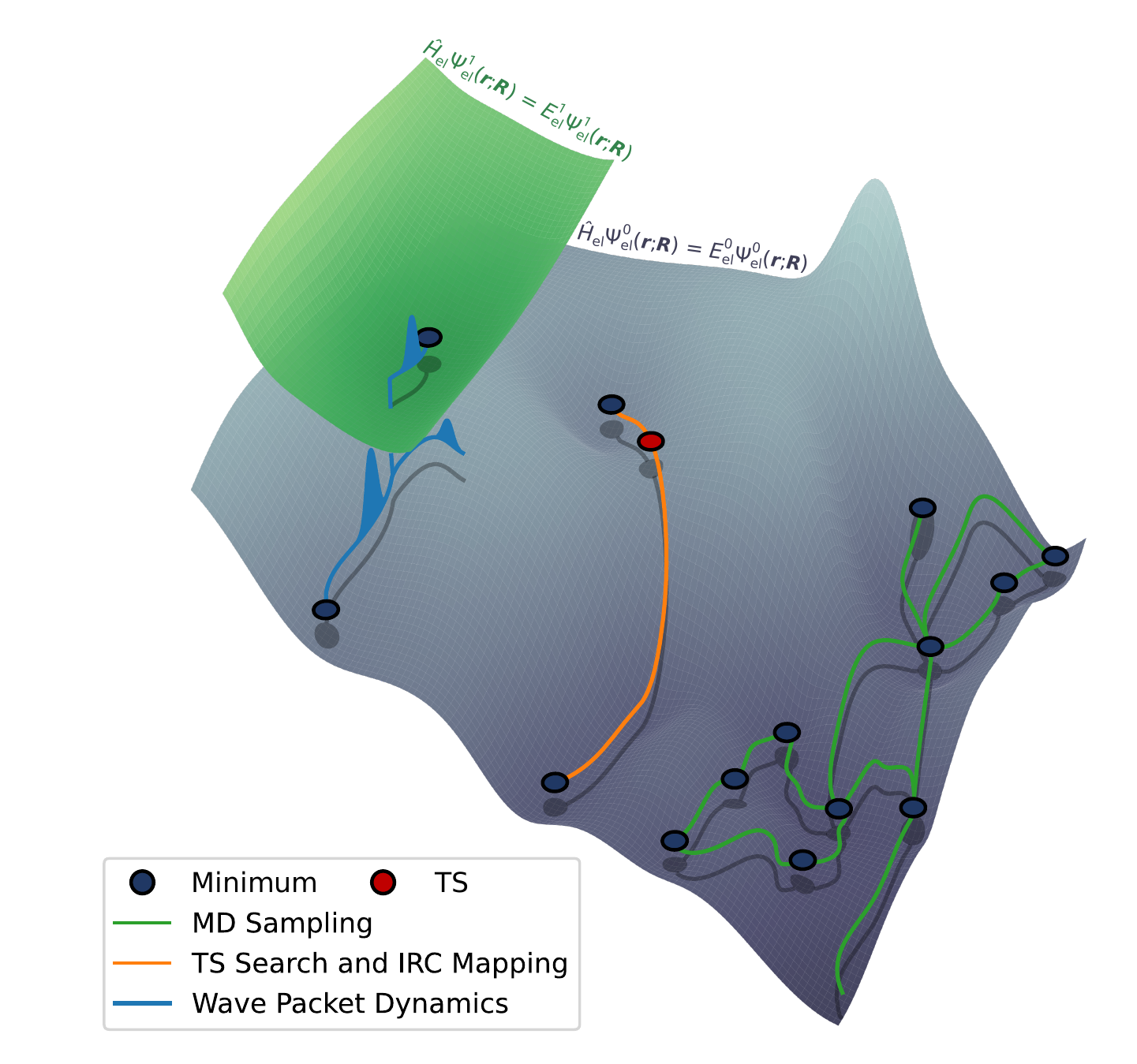}
    \caption{Schematic representation of a ground state (blue) and an excited state (green) potential energy surface, with associated electronic Schr\"odinger eigenvalue equations, on which reaction paths are explored by the three approaches discussed in this work (see legend); acronyms: MD = molecular dynamics; TS = transition state; IRC = intrinsic reaction coordinate.
}
    \label{fig:explorationOverview}
\end{figure}

An exploration of chemical reaction pathways typically starts in a local
minimum of the Born--Oppenheimer potential energy surface of a reactive system.
The dimension of this surface grows linearly with the number of atoms and
escaping the minimum in a direction that leads to a viable reaction pathway
is difficult. In this work, we first discuss stationary approaches
that assess potential exit routes toward new local minima defining new 
(meta)stable structures and then turn to their refinement. Afterwards, classical and quantum
dynamical approaches designed for the same purpose are considered 
(Figure (\ref{fig:explorationOverview}) presents a sketch of this general setting and
an overview of these different approaches).

\section{Crawling Out of Local Minima}
\label{sec:crawling}

\subsection{First-Principles Heuristics}
\label{sec:firstprinciples}
Algorithms for the automated generation of reaction coordinates are required for large-scale automated screening
of reaction paths.
A common approach is to construct such coordinates --- or, depending on the algorithm, the resulting intermediates ---
from connections between atoms (or groups of atoms) of the involved reactants, to induce the formation or 
dissociation of bonds between them\cite{Broadbelt1994, Maeda2010, Zimmerman2013, Bergeler2015, Habershon2015, Suleimanov2015, Simm2017}.
However, even if only reactions with reactive sites composed of one atom each are considered, which by far does not cover all types of reactions, their number roughly grows quadratically with the number of atoms.
Therefore, the number of potentially important reaction coordinates increases steeply with system size, 
and hence, the associated computational cost for screening all will be substantial.
However, it is not clear \textit{a priori} how many of these paths will lead to viable reactions.  

One solution to tame this combinatorial explosion of the number of possible reaction paths has been 
weeding based on chemical rules and concepts (see, \textit{e.g.}, Refs.~\citenum{Rappoport2014, Ismail2019}). 
However, such rules are usually limited in their range of applicability. 
For this reason, we have put forward the
idea of first-principles heuristics\cite{Bergeler2015}, \textit{i.e.}, harnessing concepts rooted in quantum
mechanics \cite{Grimmel2021} that can
be calculated from the electronic structure of the reactants and that are therefore applicable to any molecular
system. Hence, first-principles heuristics is agnostic with respect to the atoms or type of compounds under 
consideration. 

\begin{figure}[H]
    \centering
    \includegraphics[width=1.0\linewidth]{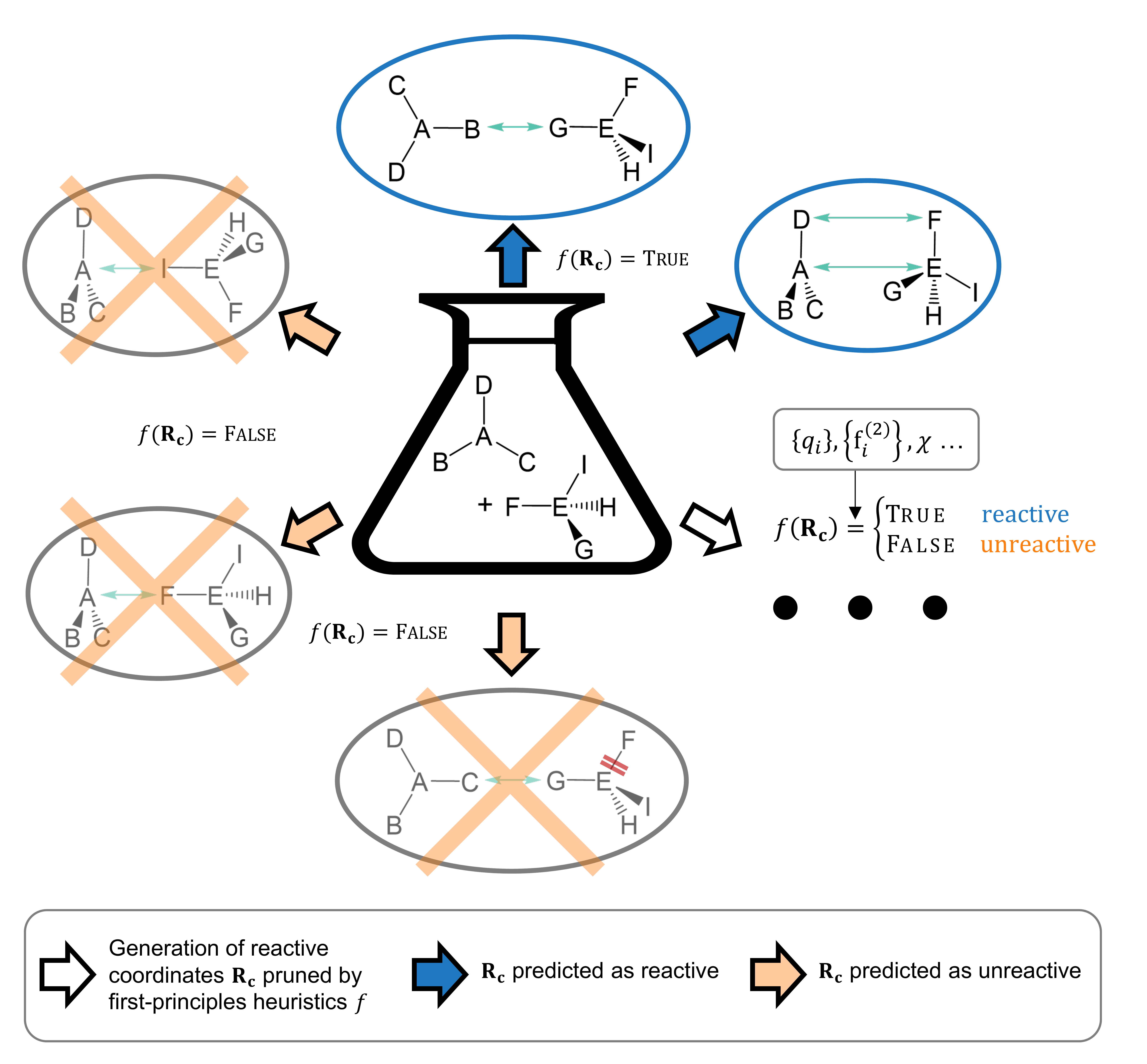}
    \caption{First-principles heuristics $f$ tames the combinatorial explosion  of potentially
reactive coordinates $\mathbf{R_C}$ based on the electronic structure of the reactants by exploiting 
a concert of chemical concepts such as partial charges $\left\{q_i\right\}$, dual descriptor 
$\left\{\text{f}^{(2)}_i\right\}$, and electronegativity $\chi$.}
    \label{fig:figure2}
\end{figure}

The usefulness of first-principles heuristics for reaction coordinate selection 
depends on reactivity predictions based on the local information available for the isolated reactants
in their equilibrium structures. Naturally,
uncertainties for such predictions can be expected as no direct information on activation barriers and 
reaction energies is available from such local descriptors \cite{Grimmel2021}.
However, as long as the reliability of the predictions is high enough to at least 
rule out a significant share of reaction coordinates that do not yield feasible reactive paths,
first-principles heuristics still tames the combinatorial explosion of the number of
reaction paths by excluding those that do not need to be inspected \cite{Grimmel2021}; see 
Figure \ref{fig:figure2}).

The idea of applying first-principles heuristics for predicting reactive sites was demonstrated 
for identifying protonation sites by analysis of the electron localization function and the 
molecular electrostatic potential \cite{Bergeler2015, Grimmel2019}.

In order to predict other types of chemical reactions, general approaches are needed based
on electronic structure theory. For example, a more diverse set of chemical concepts evaluated from 
the electronic wave function of reactants can be exploited, such as 
reactivity concepts defined within the context of conceptual density functional theory.\cite{Grimmel2021}
Instead of exclusively focusing on finding a single best reactivity descriptor, it will be important to consider 
combinations of all of them which is obvious for formal reasons, \textit{i.e.}, in view of the role that many of these concepts play
in the very same Taylor series expansion \cite{Grimmel2021}.

\subsection{Enforcing Chemical Reactions}  
\label{sec:afir_nt}
Once promising reaction coordinates are identified, the next question is how to follow such paths and extract a guess for 
the transition state (TS) structure. In single-ended approaches, which represent the typical situation of an
exploratory ansatz in which the result of a reactive system is not known from the outset, only the reactant structure 
will be known.
One of these approaches tailored for exploratory schemes is the artificial force induced reaction 
method\cite{Maeda2010,Maeda2021},
which, as the name tells, applies an artificial force between parts of a chemical system, thereby inducing chemical reactions.
The strength of the artificial force is related to the mean force acting on two argon atoms colliding with a chosen energy.
Reaction coordinates can be selected by the choice of atoms between which the force should be applied.
Another approach is based on the growing string method \cite{Zimmerman2013b},
which, starting from the reactant, incrementally extends the reaction path while constantly examining if a possible TS is traversed.

\begin{figure}[H]
 \begin{center}
  \includegraphics[width=0.7\textwidth]{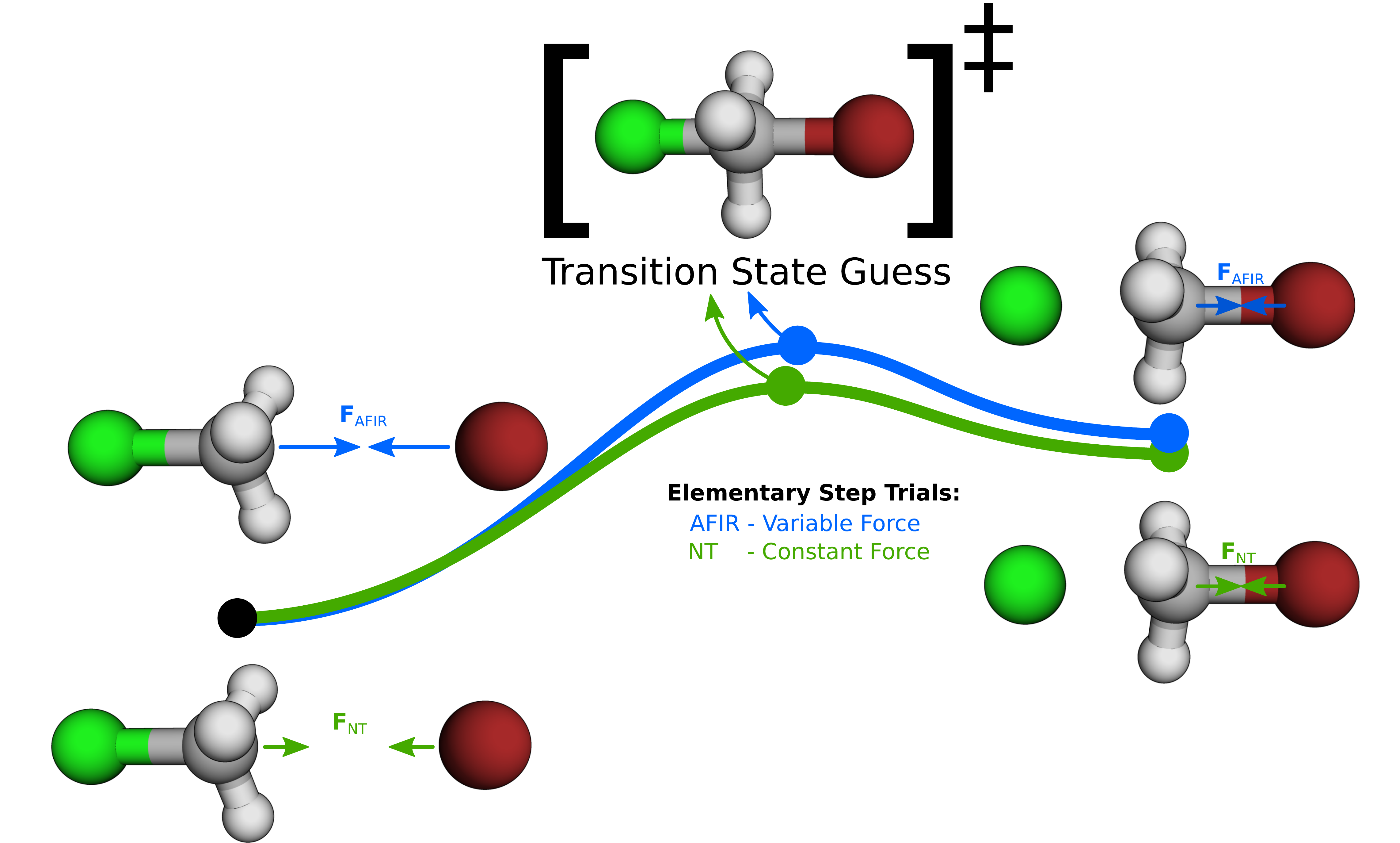}
 \end{center}
 \caption{\label{fig:figure3} Schematic representation of single-ended elementary step trials to
                  illustrate the differences in the applied force ($\textbf{F}$) in artificial force induced reaction (AFIR) and Newton trajectory (NT) algorithms.}
\end{figure}

Inspired by Newton trajectories\cite{Bofill2011}
and the recent work of Quapp {\it et al.}\cite{Quapp2020}, 
we devised for the upcoming release of our {\sc Chemoton} reaction network exploration software an algorithm
which extracts a promising TS guess structure by pushing together (or pulling apart) two predefined reactive sites with a constant force given as an input parameter.
The force parameter controls how many steps are taken until the scan stops; for instance, for colliding reactive centers a 
large parameter results in few steps and a fast screening, whereas a small parameter produces many steps and a slow screening.
Upon pushing together (or pulling apart) these reactive centers, all atoms besides these constrained atoms are continuously relaxed.
This approach allows us to start screenings for TS guess structures from anywhere on the PES, not necessarily 
starting at a minimum.
Hence, the focus of our method has been to obtain maximum energy structures as promising transition state guesses
(see Figure \ref{fig:figure3}).
A detailed description of our algorithm will be published elsewhere.
Validation of the transition state as well as refinement of the path connecting the minima is then performed in a subsequent step 
(\textit{cf.}, Section~\ref{sec:path_and_ts_refinement}).

\subsection{Interactive Quantum Mechanics}
\label{sec:iqc}
Molecules larger than about fifty atoms often feature a complex PES. An exhaustive exploration of such a PES is already very demanding
in terms of computer time. In such cases, human intuition can be efficiently utilized to steer (or limit) the exploration to selected regions of chemical reaction space or configuration space. Human 
intuition can most efficiently be harnessed in an interactive setting, \textit{i.e.}, one in which the results of a structural manipulation are promptly fed back and are therefore directly
perceivable by the person manipulating the molecule.
   
\begin{figure}[H]
\begin{center}
\includegraphics[scale=.15]{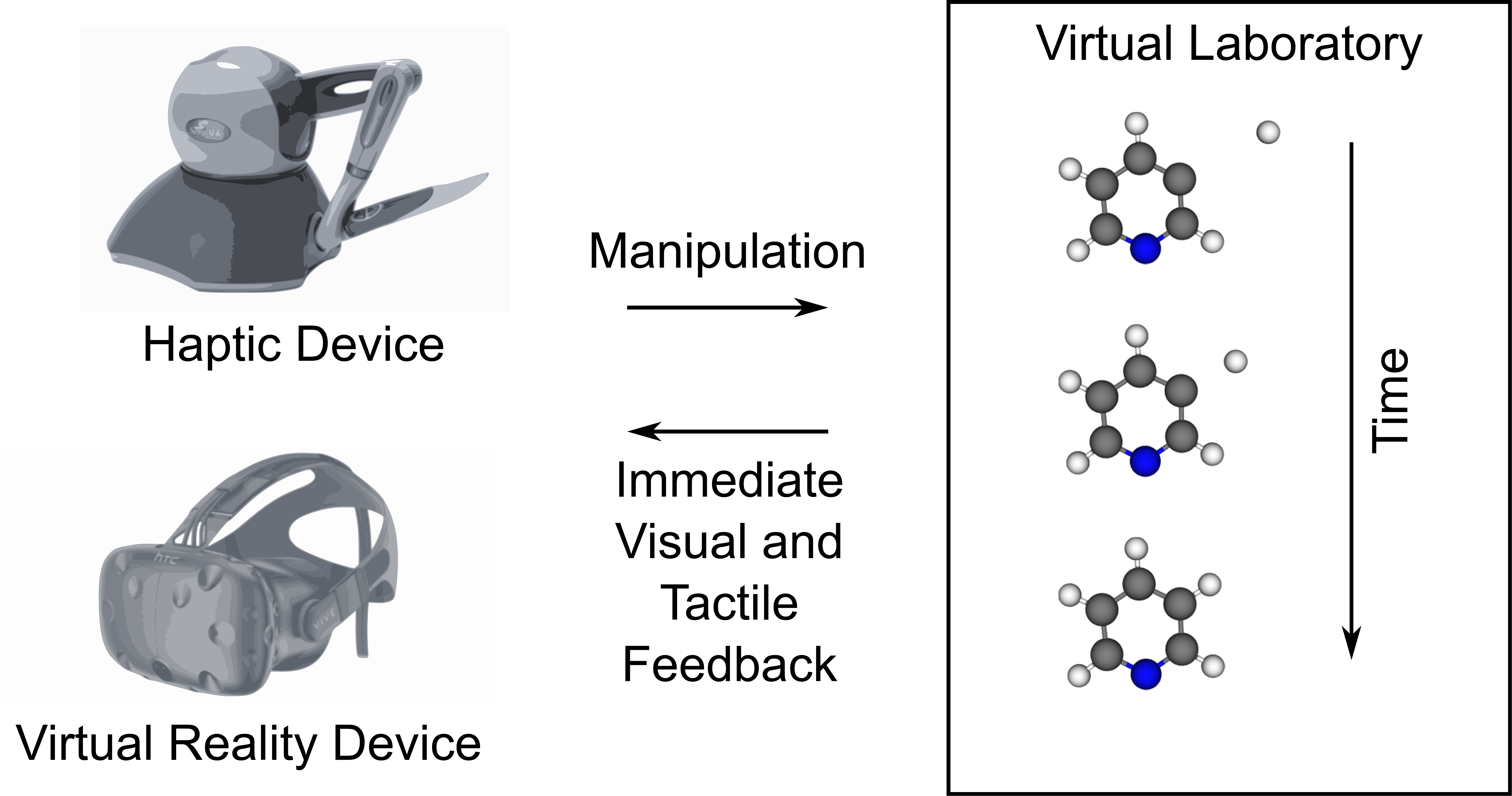}
\end{center}
\caption{\label{fig:interactive} Interactive exploration with real-time feedback allows one to take advantage of human chemical intuition. 
Two hardware extensions are shown: a haptic force feedback device (top) and a virtual reality head set (bottom).}
\end{figure}

Naturally, the bond breaking and bond forming processes of a reactive system require a description according to the laws of
quantum mechanics because only at the level of the elementary particles (\textit{i.e.}, the electrons moving in the external
field of the atomic nuclei) are all interactions well known (even exactly in the limit of an infinite speed of light).   
Then, a molecule ``reacts'' to human manipulation in a physically meaningful way. 
This then requires real-time quantum chemistry proposed by us in 2013\cite{Haag2013}, which at its core 
comprises ultra-fast quantum chemical calculations 
that deliver energies and forces almost instantaneously. 
As a result, interactive quantum mechanics allows a person to explore a PES interactively as the feedback to structural manipulations is 
directly given, \textit{i.e.}, in real time. This feedback can be visual and haptic (\textit{i.e.}, addressing our tactile sense) \cite{Marti2009,
Haag2014, Haag2014a, Weymuth2021}; see Figure \ref{fig:interactive}. 
Visual feedback is needed for an immersive experience. The degree of immersion can be enhanced by making use of virtual and/or
augmented reality technologies\cite{OConnor2018,Weymuth2021}.

\section{Refinement of Paths and Transition State Structures}
\label{sec:path_and_ts_refinement}
After an initial reaction path has been obtained from either automated or interactive methods, it needs to be refined 
in such a way that a minimum energy reaction path (MERP) is obtained. An important part of this process is the determination of the TS structure.
If the product of a reaction is unknown, a single-ended method will be necessary, which solely requires a single molecular structure as input. 
A routinely used algorithm of this class is the eigenvector following method,   
which is reliable\cite{Schlegel2003}, but also computationally expensive as every iteration requires a Hessian calculation.
The latter can be alleviated by either updating the initial Hessian based on the change of the gradient after each step with Bofill's method\cite{Bofill1994}
or by exploiting a subspace iteration scheme to approximate the decisive part of the Hessian in a systematic manner\cite{Sharada2014, Bergeler2015a}.
However, very many methods have been proposed for cases where Hessians are not available or too time-consuming to calculate.

After a transition state has been obtained, the complete MERP needs to be constructed. The simplest approach is a steepest descent optimization from the TS after an initial displacement along the lowest eigenvector of the Hessian in either direction.
If the two end points of a path are already known, for instance, when the exploration has
delivered a potentially interesting stable intermediate,
a double-ended method can be applied to obtain the TS structure, \textit{i.e.,} both the reactant(s) and the product(s) of a given chemical reaction are provided as input.
With such methods, an initial guess for the MERP can be generated from a simple linear geometric interpolation between reactant and product.
The approximate MERP is then optimized, for which most algorithms use a chain of states, \textit{i.e.}, multiple discrete structures along the path.

\begin{figure}[H]
 \begin{center}
  \includegraphics[width=0.7\textwidth]{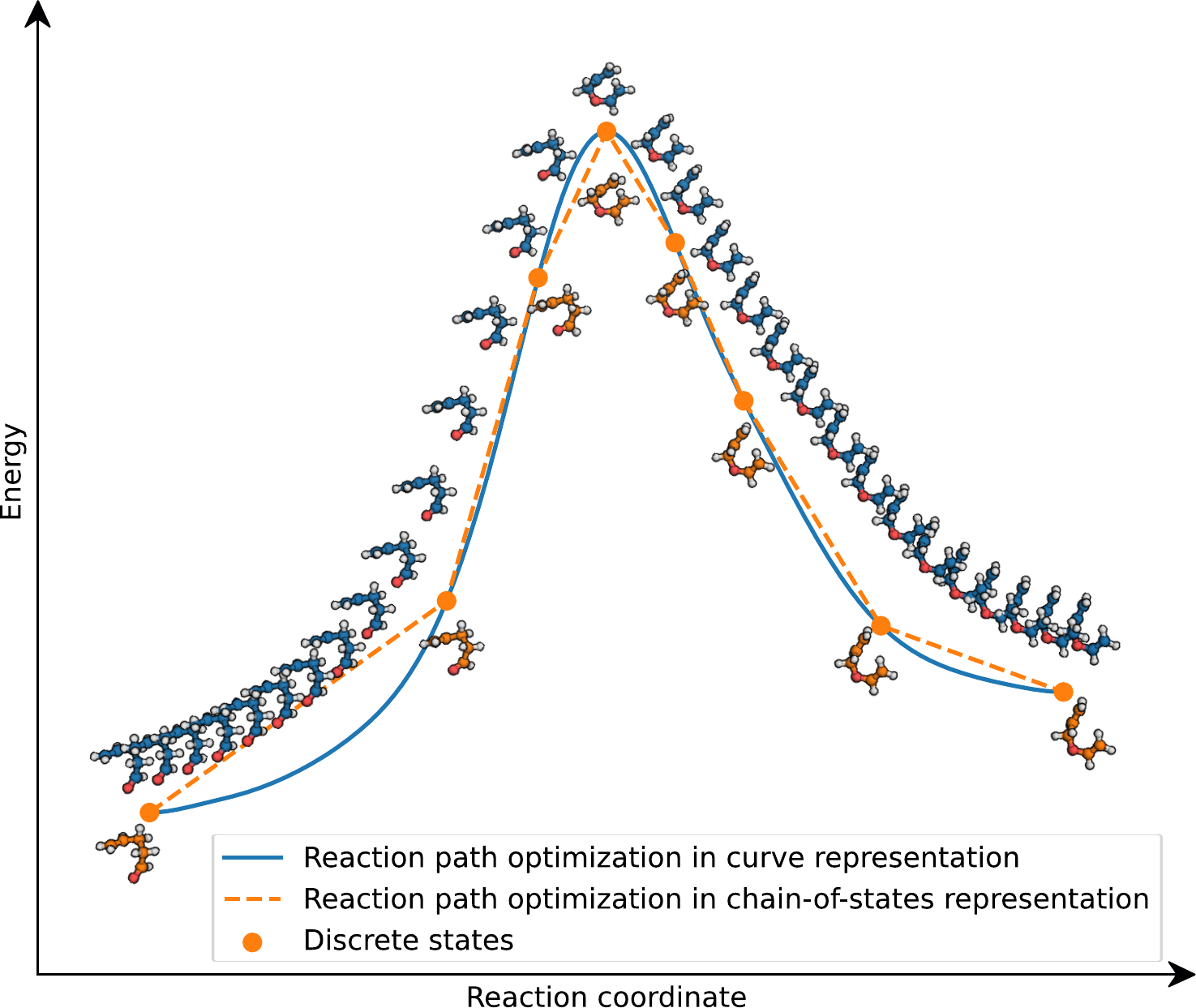}
 \end{center}
 \caption{\label{fig:figure5} 
Reaction path optimization of a Claisen rearrangement by curve optimization in blue and, for comparison,
by a connected chain of states in orange.
}
\end{figure}

By contrast, it is also possible to generate an analytical curve from the structures, which can be optimized to arrive at a truly continuous reaction path rather than at a discrete series of distinct structures 
(see Figure \ref{fig:figure5} for an example).
The splined saddle method\cite{Granot2008} applies cubic splines to optimize transition states, but it does not construct a MERP.
Another approach proposed by Birkholz and Schlegel\cite{Birkholz2015} is the variational reaction coordinate method, which minimizes the variational reaction energy.
This quantity depends on the nuclear gradients of the electronic energy, and hence, its optimization requires the electronic Hessian matrix to be repeatedly evaluated.
These approaches are related to our aforementioned curve-optimization algorithm ReaDuct\cite{Vaucher2018}.
ReaDuct directly optimizes the analytical parameters of a fitted spline curve, which only requires the electronic energy and its nuclear gradient.
Therefore, both the transition state as well as the entire MERP are obtained in a single optimization, and the resulting reaction path is truly continuous.

\section{Exploratory Dynamics Across Energy Surfaces}
\label{sec:md}
\subsection{Reactive Molecular Dynamics}
\begin{figure}[H]
 \begin{center}
  \includegraphics[width=0.8\textwidth]{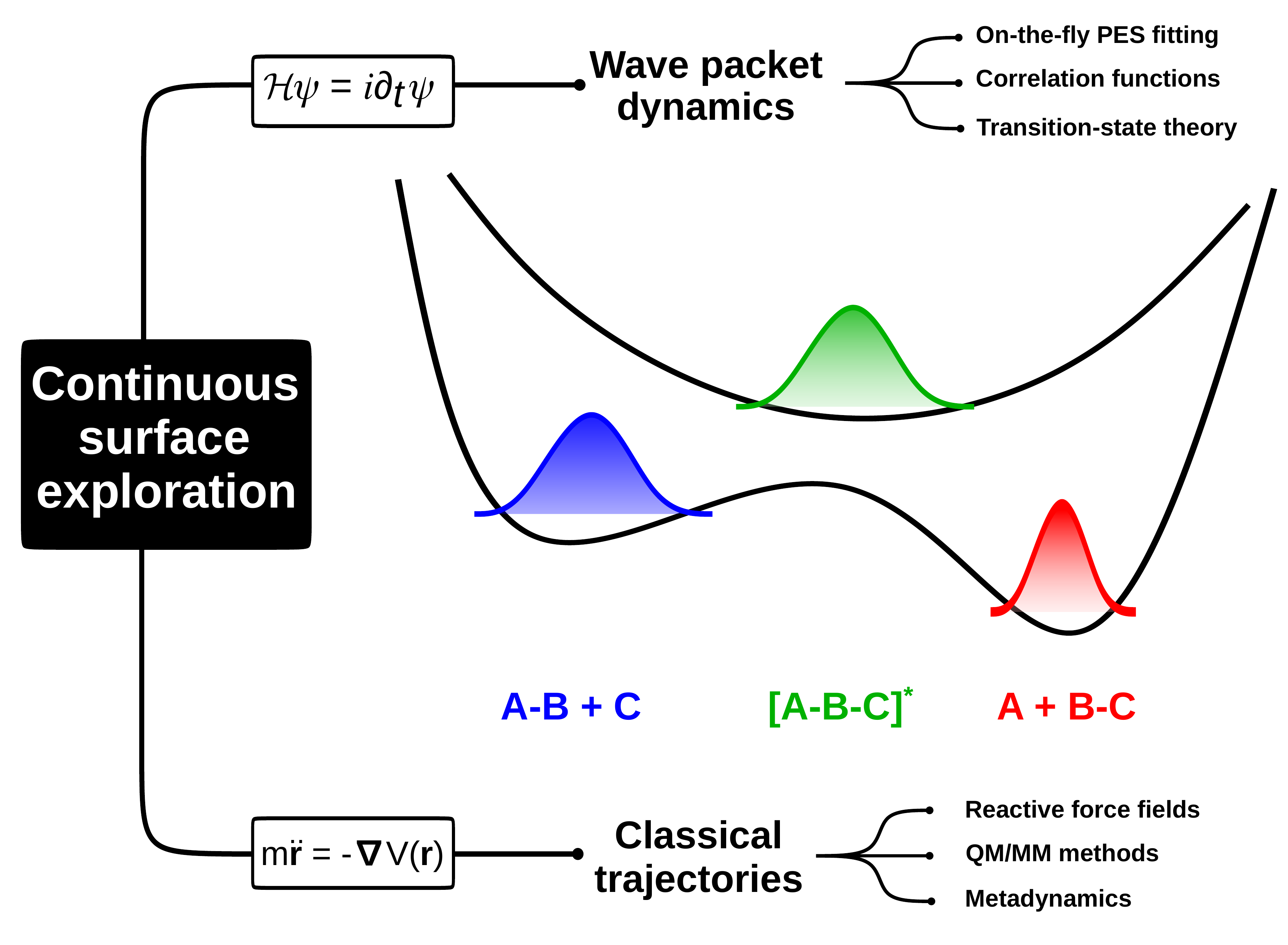}
 \end{center}
 \caption{\label{fig:figure6} 
Comparison of classical (bottom) and quantum (top) nuclear dynamics 
for exploration purposes based on trajectories and wave packet propagation, respectively. 
}
\end{figure}

Molecular dynamics (MD) simulations can directly evolve a reactive chemical system by numerical integration of Newton's equations of motion
for the nuclei or atoms, whereas the interactions are described quantum mechanically through the solution of the electronic
Schr\"odinger equation or by a reactive force field 
(see Figure \ref{fig:figure6}).

A major challenge for the MD simulation of chemical reactions is, however, their rare-event nature: 
Reaction times of complex systems can easily reach the order of milliseconds. 
However, molecular vibrations and the actual bond formation take place on picosecond time scales 
which requires MD integration time steps that are even smaller.
Hence, a huge number of MD steps is necessary to simulate a reactive system resulting in a computational effort that is unfeasible for most routine applications.
This is why enhancement techniques accelerating MD simulations are needed.

Numerous methods 
have been devised and are well established to tackle the rare-event challenge.
For instance, in metadynamics \cite{Laio2002},
which combines local elevation\cite{Huber1994} with a reduced set of collective variables,
a history-dependent bias potential is built up iteratively by adding differentiable functions (usually Gaussians) 
centered around the points in collective-variable--space that are visited. Hence, further sampling of this area is energetically
penalized and the escape from such minima is promoted.
Naturally, metadynamics schemes have been employed for the exploration of reaction paths to map out
selected regions of chemical reaction space\cite{Piccini2018, Grimme2019, Mandelli2020}.
However, the choice of appropriate collective variables is an obvious challenge with the potential to introduce a 
bias in unsupervised exploratory studies. 

If no \textit{a priori} information about a system's reactivity is given, it is often advantageous to avoid the construction of 
collective variables and instead use enhancement algorithms that do not require them.
As the probability of an energy barrier to be crossed decreases exponentially with the ratio of the barrier height over the temperature, 
one approach to accelerate reactive MD simulations is simply to increase the temperature. 
For example, for their \textit{ab initio} nanoreactor Wang \textit{et al.} combined this ansatz with a bias potential mimicking a piston that periodically compresses the system.\cite{Wang2014}
High simulation temperatures may, however, result in different reactions than those favored under ambient conditions (see, for example, Ref. \citenum{Shannon2018}).
In ``boxed molecular dynamics in energy space'' 
\cite{Shannon2018} the system's energy is only increased gradually, and hence, sampling at extremely high temperatures is avoided. 

Large nanoscale systems benefit from some sort of embedding scheme (see, \textit{e.g.}, Ref.~\citenum{Muhlbach2018} and references therein) to tame the effort for a quantum calculation
while incorporating the surrounding spectator environment in a proper way. 
This environment is often modeled with a computationally inexpensive unreactive molecular mechanics force-field, whereas the reactive center
demands energies and forces from a quantum mechanical method or a reactive force field.
Only recently, algorithms for the systematic and automated determination of a quantum mechanical region have been introduced
for such approaches\cite{Karelina2017, Brunken2021},
which are key for the easy deployment of quantum mechanical/molecular mechanics schemes in the context of vast automated explorations for
potentially important reaction paths.

\subsection{Direct Wave Packet Dynamics}
\label{sec:qd}

The molecular dynamics methods discussed above rely on two key approximations for the nuclear dynamics.
First, non-equilibrium (non-Born--Oppenheimer) electronic effects are neglected such that the nuclear rearrangement is on the ground-state PES.
Second, nuclear quantum effects are neglected.
These assumptions are certainly appropriate for studying thermal reactions in the condensed phase under non-exotic
conditions, but they are not met for photochemical processes driven by non-equilibrium electronic effects 
and for phenomena that are strongly affected by nuclear quantum effects.
``Nuclear quantum dynamics'' 
integrate the time-dependent Schr\"{o}dinger equation
(see Figure \ref{fig:figure6}) is the main route to overcome these limitations.
Naturally, the time evolution of a well-prepared wave packet inherently explores the relevant PESs.
However, explorative dynamics of this kind is typically limited by significant computational cost and the
necessity to have an analytic representation of sections of the PES available on which the
wave packet moves.
An efficient quantum dynamics scheme relies on the availability of three main building blocks: 1) an efficient method to calculate electronic energies at different, possibly many, nuclear configurations, 2) a fitting algorithm to represent the resulting energy values as a PES, and 3) a method to propagate the nuclear wave packets by integrating the time-dependent Schr\"{o}dinger equation.
Balancing the accuracy and the cost of these three components is the overarching challenge of nuclear quantum dynamics.

The multi-configurational time-dependent Hartree (MCTDH)\cite{Meyer2011} algorithm, combined with the POTFIT fitting scheme,\cite{Brommer2015} is
the canonical method for accurate vibrational quantum dynamics simulations.
As any method relying on the exact diagonalization concept, MCTDH is plagued by exponential scaling with system size.
MCTDH variants have been devised to tame this high computational cost and simulate the exact nuclear dynamics of molecules with several dozens
of fully coupled degrees of freedom.
All these methods enhance the MCTDH efficiency by applying tensor network methods to compress the wave function representation.
The most prominent example is the multi-layer MCTDH method\cite{Wang2003,Manthe2008,Burghardt2021} that relies on the hierarchical Tucker factorization of the many-body wave function.
The time-dependent density matrix renormalization group method
has been recently applied to simulate nuclear\cite{Kurashige2018,Baiardi2019,Shuai2020} 
quantum dynamics.
Density matrix renormalization group driven dynamics were shown to work for problems that are challenging for other state-of-the-art quantum dynamics schemes, which suggests that further developments may drastically enhance the efficiency of the existing MCTDH variants.

The second key component is the PES fitting algorithm.
For many applications (even when no explicit time propagation occurs), an accurate sampling of the PES is not feasible within a reasonable amount of time. 
In such cases, a PES may be constructed on the fly with general interpolation
schemes. Many such schemes have been developed over the past decades. 
One of them is interpolating moving least-squares\cite{Shepard1968}, 
which is a robust and scalable method that works with energies only (if gradients and/or Hessians are available,
however, this information can be incorporated for improved results). We applied interpolating
moving least-squares successfully in some early applications of haptic
quantum chemistry\cite{Marti2009} for reaction path exploration \cite{haag2011}, where interactive manipulation rather than explicit dynamics is the driver for
PES availability.

The development of increasingly more efficient PES fitting schemes has been powered 
by modern machine learning (ML)-based methods in recent years.
Three classes of ML algorithms have been extensively applied to obtain a compact representation
of a PES: neural networks (NN),\cite{Parrinello2007} permutationally invariant polynomials (PIP),\cite{Bowman2009} and kernel ridge regression (KRR).\cite{Dral2020}
Each of these three classes have distinctive strengths and drawbacks.
PIPs have the key advantage that they naturally encode the PES permutational symmetry upon exchange of identical atoms, 
but their optimization becomes challenging when applied to molecules with more than ten atoms.\cite{Behler2019}
Although fragmented PIPs\cite{Conte2020} have been designed to target larger systems, NNs yield more 
compact PES representation for molecules with several dozens degrees of freedom.\cite{Behler2015}
This higher representation power comes at the price that calculating the NN representation of a PES is a major challenge in practice.
KRR-based PESs represent a good compromise between PIPs and NNs since they offer a more flexible representation power and, at the same time, their optimization remains straightforward.
The efficiency of KRR fitting algorithms makes it possible to combine them with interactive quantum chemistry algorithms\cite{Amabilino2019}.

Many quantum dynamics algorithms rely, in fact, on a specific form of the PES, and therefore, they can only be combined with specific NN-based algorithms.
For instance, MCTDH requires the PES to be expressed in the ``sum-of-product'' format, which is much simpler than the forms that  
are parametrized by ML algorithms described above.
ML algorithms must, therefore, be adapted to accommodate the needs of the quantum dynamics algorithm.
For instance, Habershon and co-workers designed a specific form of the KRR algorithm that is compatible 
with the requirements of MCTDH.\cite{Habershon2019}
Similarly, neural networks with specific activation functions\cite{Manzhos2006} or topologies\cite{Koch2014} have been successfully applied to MCTDH-based quantum dynamics simulations.

These methods sacrifice NN flexibility of the PES representation to simulate accurately vibrational quantum dynamics.
The price they pay is, however, that many energy values are needed to fit the PES to balance the loss in PES flexibility.
Alternatively, it is possible to compromise the accuracy of the quantum dynamics method to maximally exploit the potential of ML algorithms.
In approximate nuclear quantum dynamics algorithms
the nuclear dynamics is driven by the local harmonic approximation of the PES.
Hence, they do not need the PES to be expressed as a global function, and for this reason, these methods can support much more flexible ML algorithms such as deep neural networks\cite{Marquetand2019} and full-dimensional kernel ridge regression models.\cite{Thiel2018}

Independent of their relative accuracy, all these methods share two main limitations with respect to quantum dynamics applications: they are often tailored to fit the PES of electronic ground states and they require a fixed set of data points to be available to calculate them.
Ideally, a propagation requires fitting an arbitrary number of excited-state PESs on-the-fly, by calculating the PES only for the nuclear configurations that are explored by the propagation.
Fulfilling these requirements is a much more complex task than the original problem of fitting high-dimensional PESs, which has already been challenging.
First, the reference energies must be calculated with high-accuracy electronic structure calculations.
Coupled cluster may be the method of choice for calculating the ground-state energy of weakly-correlated molecules, 
but such a well-defined accurate reference algorithm is missing for strongly correlated systems and, in general, for excited states.
In this respect, wave-function-based methods have witnessed an impressive development in the last decade
that made it possible to calculate very accurate energies for molecules with up to one hundred electrons;
prominent examples are tensor-based algorithms and specialized configuration interaction approaches
(see Ref. \citenum{Baiardi2020} for a discussion and for detailed references).

Even if it will be possible to routinely apply these electronic structure methods, obtaining a compact representation of the resulting PES will remain a challenging task.
Excited-state potential energy surfaces are, in fact, largely less regular than their ground-state counterparts, especially in the vicinity of conical intersections that are at the heart of photochemical processes.
The vast majority of photochemical processes are, however, determined by transitions between excited states and PESs alone are not sufficient to include these phenomena, which are driven by non-adiabatic and spin--orbit couplings.
If the choice of the PES representation may be guided by chemical intuition, this will not be true for these coupling terms that are therefore harder to represent with ML.
Early applications of ML algorithms to non-adiabatic excited-state dynamics did not attempt representing such terms with ML and calculated them on the fly in the vicinity of conical intersections.\cite{Thiel2018}
However, it has recently been shown that the representation power of deep NNs can be exploited to represent all the components that are required for accurate quantum dynamics simulations.\cite{Marquetand2019}

All new approaches discussed above demonstrate how much machine-learning methods have pushed the limits of the non-equilibrium molecular dynamics simulation methods.
However, a major open challenge has often been overlooked and this is a key issue in connection with the automatic exploration of chemical processes.
The vast majority of methods ``learn'' the PES representation before the propagation.
This is not an issue for simulating photochemical reactions where the products are known, but it becomes a severe limitation for discovering new photochemical reactivity.
On the one hand, obtaining an accurate representation of the PES for all nuclear configurations of potential relevance is unfeasible and would require an enormous number of electronic structure calculations.
On the other hand, ML-based PES fitted based on energies calculated for a partial set of nuclear configurations can lead to largely inaccurate results.
Only algorithms that can update the PES representation on the fly as the propagation proceeds can solve this issue.
The on-the-fly MCTDH variant designed by Habershon, Knowles and co-workers \cite{Habershon2019a} implements such a scheme with KRR-based algorithms.

\section{Path Information Extraction from Dynamics Data}
\subsection{State-to-State Networks}
Statically exploring a PES as described in Section~\ref{sec:crawling} automatically yields a (reaction) network with clearly defined start and end structures as well as the paths connecting these structures. However, the situation is rather different in the case of exploratory dynamics simulations. 
In the case of classical molecular dynamics, the result of the exploration are trajectories without clearly defined start and end structures. Rather, such structures need to be extracted from trajectories in a second step.

A first possibility is to manually identify structures and associated reactions.
Such an approach is very laborious and time consuming. Therefore, many different automatic approaches have been devised.
Before the advent of reactive force fields, the focus was on determining individual conformations, \textit{i.e.}, assigning very similar structures to the same conformation. This was typically done by means of a suitable clustering algorithm. A wide range of clustering algorithms to decompose the conformational space into several distinct ``bins'' has become available (many of them cluster sets of coordinates according to structural or kinetic similarity).
In order to detect reactive events such as the breaking of a bond, a different approach is usually taken. For every frame of a trajectory, the atomic connectivities are computed.
Changes in the connectivity are then interpreted as reactions\cite{Wang2014, Martinez2015, Dontgen2015, Vazquez2018, Wu2019}. 
A related approach monitors changes in the bond order matrix\cite{Liu2014, Hutchings2020, Hirai2021}.

\subsection{Transition Probability Information}

Within a classical picture, the key quantity for characterizing the reaction dynamics is the reaction constant.
The link between wave function propagation and reaction rate is the flux-flux correlation function\cite{Miller1983}.
For accurate calculations on few-particle systems, the MCTDH algorithm is the method of choice for evaluating this function.
However, the accuracy of full quantum approaches is often not needed for many quantum-chemical applications.

Quantum effects affect chemical reactions in two ways.
First, the zero-point vibrational energy (ZPVE) affects the relative energy of reactants, products, and transition states.
Second, nuclear quantum effects such as tunneling deflect the nuclear dynamics from the classical path.
While the first component influences any chemical reaction, the second one is relevant only for the dynamics of light atoms.
For reactions involving heavy atoms, a significantly cheaper alternative to exact quantum dynamics approaches is offered by transition state theory.
The classical formulation of transition state theory leads to the well-known Eyring equation.
	
Transition state theory effectively includes quantum effects associated with ZPVE effects, but it approximates tunneling effects.
In fact, it assumes that these effects are relevant only for a single, strongly anharmonic coordinate, usually corresponding to the reaction coordinate.
For this reason, transition state theory yields an inaccurate description of tunneling phenomena that are inherently multidimensional, happening under the ``deep tunneling'' regime.
Semiclassical quantum dynamics methods provide a reasonable compromise between exact quantum dynamics methods and the 
transition state theory for studying these more challenging reactions.
In this respect, the link between semiclassical nuclear dynamics and the calculation of reaction rates is the instanton theory 
which, combined with the ring polymer molecular dynamics algorithm, is currently emerging as a very powerful framework for calculating rate constants 
of reactions that are strongly affected by nuclear quantum effects.\cite{Richardson2018}

\section{Conclusions and Outlook}
Recent developments of numerical methods have pushed the range of
applicability of first-principles methods to a point where it will
become increasingly easier to apply them in an exploratory fashion
on a routine basis. This will facilitate two key developments for
molecular simulations: 1) A much broader range of situations can
be handled which will reduce the risk of overlooking potentially
important sections of the potential energy surface and hence of the
reactive dynamics of some molecular
system. 2) Unexpected discoveries can be made as the bias due to 
preexisting knowledge about a specific system is reduced to a large
extent --- up to the point where hardly any preexisting knowledge will 
be required to conduct explorations in an open-ended way.
It can be expected that work along these lines will continue and
deployment of the developed algorithms will benefit from 
software that is characterized by intuitive human--machine
interfaces requiring as little training as possible, hence
lowering the entrance barrier for researchers and non-experts
to a large extent. This will
be made viable by algorithms that act and interact as autonomously 
as possible.

\section*{Acknowledgments}
\label{sec:acknowledgments}

This work was generously supported by the Swiss National Science Foundation (SNSF) through project no.~200021\_182400.
and as part of NCCR Catalysis, a National Centre of Competence in Research funded by the SNSF.
Support from ETH Zurich is gratefully acknowledged through grant ETH-44~20-1.
MS gratefully acknowledges a Swiss Government Excellence Scholarship for Foreign Scholars and Artists.
JPU is grateful for financial support by the Deutsche Forschungsgemeinschaft (UN~417/1-1).

\providecommand{\latin}[1]{#1}
\makeatletter
\providecommand{\doi}
  {\begingroup\let\do\@makeother\dospecials
  \catcode`\{=1 \catcode`\}=2 \doi@aux}
\providecommand{\doi@aux}[1]{\endgroup\texttt{#1}}
\makeatother
\providecommand*\mcitethebibliography{\thebibliography}
\csname @ifundefined\endcsname{endmcitethebibliography}
  {\let\endmcitethebibliography\endthebibliography}{}

%
%


\begin{mcitethebibliography}{76}
\providecommand*\natexlab[1]{#1}
\providecommand*\mciteSetBstSublistMode[1]{}
\providecommand*\mciteSetBstMaxWidthForm[2]{}
\providecommand*\mciteBstWouldAddEndPuncttrue
  {\def\EndOfBibitem{\unskip.}}
\providecommand*\mciteBstWouldAddEndPunctfalse
  {\let\EndOfBibitem\relax}
\providecommand*\mciteSetBstMidEndSepPunct[3]{}
\providecommand*\mciteSetBstSublistLabelBeginEnd[3]{}
\providecommand*\EndOfBibitem{}
\mciteSetBstSublistMode{f}
\mciteSetBstMaxWidthForm{subitem}{(\alph{mcitesubitemcount})}
\mciteSetBstSublistLabelBeginEnd
  {\mcitemaxwidthsubitemform\space}
  {\relax}
  {\relax}

\bibitem[Haag and Reiher(2013)Haag, and Reiher]{Haag2013}
Haag,~M.~P.; Reiher,~M. {Real-time Quantum Chemistry}. \emph{Int. J. Quantum
  Chem.} \textbf{2013}, \emph{113}, 8--20\relax
\mciteBstWouldAddEndPuncttrue
\mciteSetBstMidEndSepPunct{\mcitedefaultmidpunct}
{\mcitedefaultendpunct}{\mcitedefaultseppunct}\relax
\EndOfBibitem
\bibitem[Bergeler \latin{et~al.}(2015)Bergeler, Simm, Proppe, and
  Reiher]{Bergeler2015}
Bergeler,~M.; Simm,~G.~N.; Proppe,~J.; Reiher,~M. {Heuristics-Guided
  Exploration of Reaction Mechanisms}. \emph{J. Chem. Theory Comput.}
  \textbf{2015}, \emph{11}, 5712--5722\relax
\mciteBstWouldAddEndPuncttrue
\mciteSetBstMidEndSepPunct{\mcitedefaultmidpunct}
{\mcitedefaultendpunct}{\mcitedefaultseppunct}\relax
\EndOfBibitem
\bibitem[Vaucher and Reiher(2018)Vaucher, and Reiher]{Vaucher2018}
Vaucher,~A.~C.; Reiher,~M. {Minimum Energy Paths and Transition States by Curve
  Optimization}. \emph{J. Chem. Theory Comput.} \textbf{2018}, \emph{14},
  3091--3099\relax
\mciteBstWouldAddEndPuncttrue
\mciteSetBstMidEndSepPunct{\mcitedefaultmidpunct}
{\mcitedefaultendpunct}{\mcitedefaultseppunct}\relax
\EndOfBibitem
\bibitem[Baiardi and Reiher(2019)Baiardi, and Reiher]{Baiardi2019}
Baiardi,~A.; Reiher,~M. {Large-scale quantum-dynamics with matrix product
  states}. \emph{J. Chem. Theory Comput.} \textbf{2019}, \emph{15},
  3481--3498\relax
\mciteBstWouldAddEndPuncttrue
\mciteSetBstMidEndSepPunct{\mcitedefaultmidpunct}
{\mcitedefaultendpunct}{\mcitedefaultseppunct}\relax
\EndOfBibitem
\bibitem[Bofill and Quapp(2020)Bofill, and Quapp]{Bofill2020}
Bofill,~J.~M.; Quapp,~W. {Calculus of Variations as a Basic Tool for Modelling
  of Reaction Paths and Localisation of Stationary Points on Potential Energy
  Surfaces}. \emph{Mol. Phys.} \textbf{2020}, \emph{118}, e1667035\relax
\mciteBstWouldAddEndPuncttrue
\mciteSetBstMidEndSepPunct{\mcitedefaultmidpunct}
{\mcitedefaultendpunct}{\mcitedefaultseppunct}\relax
\EndOfBibitem
\bibitem[V{\'a}zquez \latin{et~al.}(2018)V{\'a}zquez, Otero, and
  Mart{\'i}nez-N{\'u}{\~n}ez]{Vazquez2018}
V{\'a}zquez,~S.~A.; Otero,~X.~L.; Mart{\'i}nez-N{\'u}{\~n}ez,~E. {A
  Trajectory-Based Method to Explore Reaction Mechanisms}. \emph{Molecules}
  \textbf{2018}, \emph{23}, 3156\relax
\mciteBstWouldAddEndPuncttrue
\mciteSetBstMidEndSepPunct{\mcitedefaultmidpunct}
{\mcitedefaultendpunct}{\mcitedefaultseppunct}\relax
\EndOfBibitem
\bibitem[Dewyer \latin{et~al.}(2018)Dewyer, Arg\"{u}elles, and
  Zimmerman]{Dewyer2018}
Dewyer,~A.~L.; Arg\"{u}elles,~A.~J.; Zimmerman,~P.~M. {Methods for exploring
  reaction space in molecular systems}. \emph{WIREs Comput. Mol. Sci.}
  \textbf{2018}, \emph{8}, e1354\relax
\mciteBstWouldAddEndPuncttrue
\mciteSetBstMidEndSepPunct{\mcitedefaultmidpunct}
{\mcitedefaultendpunct}{\mcitedefaultseppunct}\relax
\EndOfBibitem
\bibitem[Simm \latin{et~al.}(2019)Simm, Vaucher, and Reiher]{Simm2019}
Simm,~G.~N.; Vaucher,~A.~C.; Reiher,~M. {Exploration of Reaction Pathways and
  Chemical Transformation Networks}. \emph{J. Phys. Chem. A} \textbf{2019},
  \emph{123}, 385--399\relax
\mciteBstWouldAddEndPuncttrue
\mciteSetBstMidEndSepPunct{\mcitedefaultmidpunct}
{\mcitedefaultendpunct}{\mcitedefaultseppunct}\relax
\EndOfBibitem
\bibitem[Unsleber and Reiher(2020)Unsleber, and Reiher]{Unsleber2020}
Unsleber,~J.~P.; Reiher,~M. {The Exploration of Chemical Reaction Networks}.
  \emph{Annu. Rev. Phys. Chem.} \textbf{2020}, \emph{71}, 121--142\relax
\mciteBstWouldAddEndPuncttrue
\mciteSetBstMidEndSepPunct{\mcitedefaultmidpunct}
{\mcitedefaultendpunct}{\mcitedefaultseppunct}\relax
\EndOfBibitem
\bibitem[Maeda and Harabuchi(2021)Maeda, and Harabuchi]{Maeda2021}
Maeda,~S.; Harabuchi,~Y. {Exploring Paths of Chemical Transformations in
  Molecular and Periodic Systems: An Approach Utilizing Force}. \emph{WIREs
  Comput. Mol. Sci.} \textbf{2021}, \emph{11}, e1538\relax
\mciteBstWouldAddEndPuncttrue
\mciteSetBstMidEndSepPunct{\mcitedefaultmidpunct}
{\mcitedefaultendpunct}{\mcitedefaultseppunct}\relax
\EndOfBibitem
\bibitem[Broadbelt \latin{et~al.}(1994)Broadbelt, Stark, and
  Klein]{Broadbelt1994}
Broadbelt,~L.~J.; Stark,~S.~M.; Klein,~M.~T. {Computer Generated Pyrolysis
  Modeling: On-the-fly Generation of Species, Reactions, and Rates}. \emph{Ind.
  Eng. Chem. Res.} \textbf{1994}, \emph{33}, 790--799\relax
\mciteBstWouldAddEndPuncttrue
\mciteSetBstMidEndSepPunct{\mcitedefaultmidpunct}
{\mcitedefaultendpunct}{\mcitedefaultseppunct}\relax
\EndOfBibitem
\bibitem[Maeda and Morokuma(2010)Maeda, and Morokuma]{Maeda2010}
Maeda,~S.; Morokuma,~K. {Communications: A Systematic Method for Locating
  Transition Structures of A + B $\rightarrow$ X Type Reactions}. \emph{J.
  Chem. Phys.} \textbf{2010}, \emph{132}, 241102\relax
\mciteBstWouldAddEndPuncttrue
\mciteSetBstMidEndSepPunct{\mcitedefaultmidpunct}
{\mcitedefaultendpunct}{\mcitedefaultseppunct}\relax
\EndOfBibitem
\bibitem[Zimmerman(2013)]{Zimmerman2013}
Zimmerman,~P.~M. {Automated Discovery of Chemically Reasonable Elementary
  Reaction Steps}. \emph{J. Comput. Chem.} \textbf{2013}, \emph{34},
  1385--1392\relax
\mciteBstWouldAddEndPuncttrue
\mciteSetBstMidEndSepPunct{\mcitedefaultmidpunct}
{\mcitedefaultendpunct}{\mcitedefaultseppunct}\relax
\EndOfBibitem
\bibitem[Habershon(2015)]{Habershon2015}
Habershon,~S. {Sampling Reactive Pathways with Random Walks in Chemical Space:
  Applications to Molecular Dissociation and Catalysis}. \emph{J. Chem. Phys.}
  \textbf{2015}, \emph{143}, 094106\relax
\mciteBstWouldAddEndPuncttrue
\mciteSetBstMidEndSepPunct{\mcitedefaultmidpunct}
{\mcitedefaultendpunct}{\mcitedefaultseppunct}\relax
\EndOfBibitem
\bibitem[Suleimanov and Green(2015)Suleimanov, and Green]{Suleimanov2015}
Suleimanov,~Y.~V.; Green,~W.~H. {Automated Discovery of Elementary Chemical
  Reaction Steps Using Freezing String and Berny Optimization Methods}.
  \emph{J. Chem. Theory Comput.} \textbf{2015}, \emph{11}, 4248--4259\relax
\mciteBstWouldAddEndPuncttrue
\mciteSetBstMidEndSepPunct{\mcitedefaultmidpunct}
{\mcitedefaultendpunct}{\mcitedefaultseppunct}\relax
\EndOfBibitem
\bibitem[Simm and Reiher(2017)Simm, and Reiher]{Simm2017}
Simm,~G.~N.; Reiher,~M. {Context-Driven Exploration of Complex Chemical
  Reaction Networks}. \emph{J. Chem. Theory Comput.} \textbf{2017}, \emph{13},
  6108--6119\relax
\mciteBstWouldAddEndPuncttrue
\mciteSetBstMidEndSepPunct{\mcitedefaultmidpunct}
{\mcitedefaultendpunct}{\mcitedefaultseppunct}\relax
\EndOfBibitem
\bibitem[Rappoport and Aspuru-Guzik(2014)Rappoport, and
  Aspuru-Guzik]{Rappoport2014}
Rappoport,~D.; Aspuru-Guzik,~A. {Complex Chemical Reaction Networks from
  Heuristics-Aided Quantum Chemistry}. \emph{J. Chem. Theory Comput.}
  \textbf{2014}, \emph{10}, 897--907\relax
\mciteBstWouldAddEndPuncttrue
\mciteSetBstMidEndSepPunct{\mcitedefaultmidpunct}
{\mcitedefaultendpunct}{\mcitedefaultseppunct}\relax
\EndOfBibitem
\bibitem[Ismail \latin{et~al.}(2019)Ismail, Stuttaford-Fowler, Ashok,
  Robertson, and Habershon]{Ismail2019}
Ismail,~I.; Stuttaford-Fowler,~H. B. V.~A.; Ashok,~C.~O.; Robertson,~C.;
  Habershon,~S. {Automatic Proposal of Multistep Reaction Mechanisms using a
  Graph-Driven Search}. \emph{J. Phys. Chem. A} \textbf{2019}, \emph{123},
  3407--3417\relax
\mciteBstWouldAddEndPuncttrue
\mciteSetBstMidEndSepPunct{\mcitedefaultmidpunct}
{\mcitedefaultendpunct}{\mcitedefaultseppunct}\relax
\EndOfBibitem
\bibitem[Grimmel and Reiher(2021)Grimmel, and Reiher]{Grimmel2021}
Grimmel,~S.~A.; Reiher,~M. {On the Predictive Power of Chemical Concepts}.
  \emph{Chimia} \textbf{2021}, \emph{75}, 311--318\relax
\mciteBstWouldAddEndPuncttrue
\mciteSetBstMidEndSepPunct{\mcitedefaultmidpunct}
{\mcitedefaultendpunct}{\mcitedefaultseppunct}\relax
\EndOfBibitem
\bibitem[Grimmel and Reiher(2019)Grimmel, and Reiher]{Grimmel2019}
Grimmel,~S.~A.; Reiher,~M. {The Electrostatic Potential as a Descriptor for the
  Protonation Propensity in Automated Exploration of Reaction Mechanisms}.
  \emph{Faraday Discuss.} \textbf{2019}, \emph{220}, 443--463\relax
\mciteBstWouldAddEndPuncttrue
\mciteSetBstMidEndSepPunct{\mcitedefaultmidpunct}
{\mcitedefaultendpunct}{\mcitedefaultseppunct}\relax
\EndOfBibitem
\bibitem[Zimmerman(2013)]{Zimmerman2013b}
Zimmerman,~P. {Reliable Transition State Searches Integrated with the Growing
  String Method}. \emph{J. Chem. Theory Comput.} \textbf{2013}, \emph{9},
  3043--3050\relax
\mciteBstWouldAddEndPuncttrue
\mciteSetBstMidEndSepPunct{\mcitedefaultmidpunct}
{\mcitedefaultendpunct}{\mcitedefaultseppunct}\relax
\EndOfBibitem
\bibitem[Bofill and Quapp(2011)Bofill, and Quapp]{Bofill2011}
Bofill,~J.~M.; Quapp,~W. {Variational Nature, Integration, and Properties of
  Newton Reaction Path}. \emph{J. Chem. Phys.} \textbf{2011}, \emph{134},
  074101\relax
\mciteBstWouldAddEndPuncttrue
\mciteSetBstMidEndSepPunct{\mcitedefaultmidpunct}
{\mcitedefaultendpunct}{\mcitedefaultseppunct}\relax
\EndOfBibitem
\bibitem[Quapp and Bofill(2020)Quapp, and Bofill]{Quapp2020}
Quapp,~W.; Bofill,~J.~M. {Some Mathematical Reasoning on the Artificial Force
  Induced Reaction Method}. \emph{J. Comput. Chem.} \textbf{2020}, \emph{41},
  629--634\relax
\mciteBstWouldAddEndPuncttrue
\mciteSetBstMidEndSepPunct{\mcitedefaultmidpunct}
{\mcitedefaultendpunct}{\mcitedefaultseppunct}\relax
\EndOfBibitem
\bibitem[Marti and Reiher(2009)Marti, and Reiher]{Marti2009}
Marti,~K.~H.; Reiher,~M. {Haptic Quantum Chemistry}. \emph{J. Comput. Chem.}
  \textbf{2009}, \emph{30}, 2010--2020\relax
\mciteBstWouldAddEndPuncttrue
\mciteSetBstMidEndSepPunct{\mcitedefaultmidpunct}
{\mcitedefaultendpunct}{\mcitedefaultseppunct}\relax
\EndOfBibitem
\bibitem[Haag and Reiher(2014)Haag, and Reiher]{Haag2014}
Haag,~M.~P.; Reiher,~M. {Studying Chemical Reactivity in a Virtual
  Environment}. \emph{Faraday Discuss.} \textbf{2014}, \emph{169},
  89--118\relax
\mciteBstWouldAddEndPuncttrue
\mciteSetBstMidEndSepPunct{\mcitedefaultmidpunct}
{\mcitedefaultendpunct}{\mcitedefaultseppunct}\relax
\EndOfBibitem
\bibitem[Haag \latin{et~al.}(2014)Haag, Vaucher, Bosson, Redon, and
  Reiher]{Haag2014a}
Haag,~M.~P.; Vaucher,~A.~C.; Bosson,~M.; Redon,~S.; Reiher,~M. {Interactive
  Chemical Reactivity Exploration}. \emph{ChemPhysChem} \textbf{2014},
  \emph{15}, 3301--3319\relax
\mciteBstWouldAddEndPuncttrue
\mciteSetBstMidEndSepPunct{\mcitedefaultmidpunct}
{\mcitedefaultendpunct}{\mcitedefaultseppunct}\relax
\EndOfBibitem
\bibitem[Weymuth and Reiher(2021)Weymuth, and Reiher]{Weymuth2021}
Weymuth,~T.; Reiher,~M. {Immersive Interactive Quantum Mechanics for Teaching
  and Learning Chemistry}. \emph{Chimia} \textbf{2021}, \emph{75}, 45--49\relax
\mciteBstWouldAddEndPuncttrue
\mciteSetBstMidEndSepPunct{\mcitedefaultmidpunct}
{\mcitedefaultendpunct}{\mcitedefaultseppunct}\relax
\EndOfBibitem
\bibitem[O'Connor \latin{et~al.}(2018)O'Connor, Deeks, Dawn, Metatla, Roudaut,
  Sutton, Thomas, Glowacki, Sage, Tew, Wonnacott, Bates, Mulholland, and
  Glowacki]{OConnor2018}
O'Connor,~M.; Deeks,~H.~M.; Dawn,~E.; Metatla,~O.; Roudaut,~A.; Sutton,~M.;
  Thomas,~L.~M.; Glowacki,~B.~R.; Sage,~R.; Tew,~P.; Wonnacott,~M.; Bates,~P.;
  Mulholland,~A.~J.; Glowacki,~D.~R. {Sampling Molecular Conformations and
  Dynamics in a Multiuser Virtual Reality Framework}. \emph{Sci. Adv.}
  \textbf{2018}, \emph{4}, eaat2731\relax
\mciteBstWouldAddEndPuncttrue
\mciteSetBstMidEndSepPunct{\mcitedefaultmidpunct}
{\mcitedefaultendpunct}{\mcitedefaultseppunct}\relax
\EndOfBibitem
\bibitem[Schlegel(2003)]{Schlegel2003}
Schlegel,~H.~B. {Exploring potential energy surfaces for chemical reactions: An
  overview of some practical methods}. \emph{J. Comput. Chem.} \textbf{2003},
  \emph{24}, 1514--1527\relax
\mciteBstWouldAddEndPuncttrue
\mciteSetBstMidEndSepPunct{\mcitedefaultmidpunct}
{\mcitedefaultendpunct}{\mcitedefaultseppunct}\relax
\EndOfBibitem
\bibitem[Bofill(1994)]{Bofill1994}
Bofill,~J.~M. {Updated Hessian matrix and the restricted step method for
  locating transition structures}. \emph{J. Comput. Chem.} \textbf{1994},
  \emph{15}, 1--11\relax
\mciteBstWouldAddEndPuncttrue
\mciteSetBstMidEndSepPunct{\mcitedefaultmidpunct}
{\mcitedefaultendpunct}{\mcitedefaultseppunct}\relax
\EndOfBibitem
\bibitem[Sharada \latin{et~al.}(2014)Sharada, Bell, and
  Head-Gordon]{Sharada2014}
Sharada,~S.~M.; Bell,~A.~T.; Head-Gordon,~M. {A finite difference Davidson
  procedure to sidestep full \textit{ab initio} hessian calculation:
  Application to characterization of stationary points and transition state
  searches}. \emph{J. Chme. Phys.} \textbf{2014}, \emph{140}, 164115\relax
\mciteBstWouldAddEndPuncttrue
\mciteSetBstMidEndSepPunct{\mcitedefaultmidpunct}
{\mcitedefaultendpunct}{\mcitedefaultseppunct}\relax
\EndOfBibitem
\bibitem[Bergeler \latin{et~al.}(2015)Bergeler, Hermann, and
  Reiher]{Bergeler2015a}
Bergeler,~M.; Hermann,~C.; Reiher,~M. {Mode-tracking based stationary-point
  optimization}. \emph{J. Comput. Chem.} \textbf{2015}, \emph{36},
  1429--1438\relax
\mciteBstWouldAddEndPuncttrue
\mciteSetBstMidEndSepPunct{\mcitedefaultmidpunct}
{\mcitedefaultendpunct}{\mcitedefaultseppunct}\relax
\EndOfBibitem
\bibitem[Granot and Baer(2008)Granot, and Baer]{Granot2008}
Granot,~R.; Baer,~R. {A Spline for Your Saddle}. \emph{J. Chem. Phys.}
  \textbf{2008}, \emph{128}, 184111\relax
\mciteBstWouldAddEndPuncttrue
\mciteSetBstMidEndSepPunct{\mcitedefaultmidpunct}
{\mcitedefaultendpunct}{\mcitedefaultseppunct}\relax
\EndOfBibitem
\bibitem[Birkholz and Schlegel(2015)Birkholz, and Schlegel]{Birkholz2015}
Birkholz,~A.~B.; Schlegel,~H.~B. {Path optimization by a variational reaction
  coordinate method. I. Development of formalism and algorithms}. \emph{J.
  Chem. Phys.} \textbf{2015}, \emph{143}, 244101\relax
\mciteBstWouldAddEndPuncttrue
\mciteSetBstMidEndSepPunct{\mcitedefaultmidpunct}
{\mcitedefaultendpunct}{\mcitedefaultseppunct}\relax
\EndOfBibitem
\bibitem[Laio and Parrinello(2002)Laio, and Parrinello]{Laio2002}
Laio,~A.; Parrinello,~M. {Escaping Free-Energy Minima}. \emph{Proc. Natl. Acad.
  Sci. USA} \textbf{2002}, \emph{99}, 12562--12566\relax
\mciteBstWouldAddEndPuncttrue
\mciteSetBstMidEndSepPunct{\mcitedefaultmidpunct}
{\mcitedefaultendpunct}{\mcitedefaultseppunct}\relax
\EndOfBibitem
\bibitem[Huber \latin{et~al.}(1994)Huber, Torda, and {van
  Gunsteren}]{Huber1994}
Huber,~T.; Torda,~A.~E.; {van Gunsteren},~W.~F. Local Elevation: A Method for
  Improving the Searching Properties of Molecular Dynamics Simulation. \emph{J.
  Comput.-Aided Mol. Des.} \textbf{1994}, \emph{8}, 695--708\relax
\mciteBstWouldAddEndPuncttrue
\mciteSetBstMidEndSepPunct{\mcitedefaultmidpunct}
{\mcitedefaultendpunct}{\mcitedefaultseppunct}\relax
\EndOfBibitem
\bibitem[Piccini \latin{et~al.}(2018)Piccini, Mendels, and
  Parrinello]{Piccini2018}
Piccini,~G.; Mendels,~D.; Parrinello,~M. {Metadynamics with Discriminants: A
  Tool for Understanding Chemistry}. \emph{J. Chem. Theory Comput.}
  \textbf{2018}, \emph{14}, 5040--5044\relax
\mciteBstWouldAddEndPuncttrue
\mciteSetBstMidEndSepPunct{\mcitedefaultmidpunct}
{\mcitedefaultendpunct}{\mcitedefaultseppunct}\relax
\EndOfBibitem
\bibitem[Grimme(2019)]{Grimme2019}
Grimme,~S. {Exploration of Chemical Compound, Conformer, and Reaction Space
  with Meta-Dynamics Simulations Based on Tight-Binding Quantum Chemical
  Calculations}. \emph{J. Chem. Theory Comput.} \textbf{2019}, \emph{15},
  2847--2862\relax
\mciteBstWouldAddEndPuncttrue
\mciteSetBstMidEndSepPunct{\mcitedefaultmidpunct}
{\mcitedefaultendpunct}{\mcitedefaultseppunct}\relax
\EndOfBibitem
\bibitem[Mandelli \latin{et~al.}(2020)Mandelli, Hirshberg, and
  Parrinello]{Mandelli2020}
Mandelli,~D.; Hirshberg,~B.; Parrinello,~M. {Metadynamics of Paths}.
  \emph{Phys. Rev. Lett.} \textbf{2020}, \emph{125}, 026001\relax
\mciteBstWouldAddEndPuncttrue
\mciteSetBstMidEndSepPunct{\mcitedefaultmidpunct}
{\mcitedefaultendpunct}{\mcitedefaultseppunct}\relax
\EndOfBibitem
\bibitem[Wang \latin{et~al.}(2014)Wang, Titov, McGibbon, Liu, Pande, and
  Mart{\'i}nez]{Wang2014}
Wang,~L.-P.; Titov,~A.; McGibbon,~R.; Liu,~F.; Pande,~V.~S.;
  Mart{\'i}nez,~T.~J. {{Discovering chemistry with an \textit{ab initio}
  nanoreactor}}. \emph{Nature Chem.} \textbf{2014}, \emph{6}, 1044--1048\relax
\mciteBstWouldAddEndPuncttrue
\mciteSetBstMidEndSepPunct{\mcitedefaultmidpunct}
{\mcitedefaultendpunct}{\mcitedefaultseppunct}\relax
\EndOfBibitem
\bibitem[Shannon \latin{et~al.}(2018)Shannon, Amabilino, O'Connor, Shalishilin,
  and Glowacki]{Shannon2018}
Shannon,~R.~J.; Amabilino,~S.; O'Connor,~M.; Shalishilin,~D.~V.;
  Glowacki,~D.~R. {Adaptively Accelerating Reactive Molecular Dynamics Using
  Boxed Molecular Dynamics in Energy Space}. \emph{J. Chem. Theory Comput.}
  \textbf{2018}, \emph{14}, 4541--4552\relax
\mciteBstWouldAddEndPuncttrue
\mciteSetBstMidEndSepPunct{\mcitedefaultmidpunct}
{\mcitedefaultendpunct}{\mcitedefaultseppunct}\relax
\EndOfBibitem
\bibitem[M{\"u}hlbach and Reiher(2018)M{\"u}hlbach, and Reiher]{Muhlbach2018}
M{\"u}hlbach,~A.~H.; Reiher,~M. {Quantum system partitioning at the
  single-particle level}. \emph{J. Chem. Phys.} \textbf{2018}, \emph{149},
  184104\relax
\mciteBstWouldAddEndPuncttrue
\mciteSetBstMidEndSepPunct{\mcitedefaultmidpunct}
{\mcitedefaultendpunct}{\mcitedefaultseppunct}\relax
\EndOfBibitem
\bibitem[Karelina and Kulik(2017)Karelina, and Kulik]{Karelina2017}
Karelina,~M.; Kulik,~H.~J. {Systematic Quantum Mechanical Region Determination
  in QM/MM Simulation}. \emph{J. Chem. Theory Comput.} \textbf{2017},
  \emph{13}, 563--576\relax
\mciteBstWouldAddEndPuncttrue
\mciteSetBstMidEndSepPunct{\mcitedefaultmidpunct}
{\mcitedefaultendpunct}{\mcitedefaultseppunct}\relax
\EndOfBibitem
\bibitem[Brunken and Reiher(2021)Brunken, and Reiher]{Brunken2021}
Brunken,~C.; Reiher,~M. {Automated Construction of Quantum--Classical Hybrid
  Models}. \emph{J. Chem. Theory Comput.} \textbf{2021}, \emph{17},
  3797--3813\relax
\mciteBstWouldAddEndPuncttrue
\mciteSetBstMidEndSepPunct{\mcitedefaultmidpunct}
{\mcitedefaultendpunct}{\mcitedefaultseppunct}\relax
\EndOfBibitem
\bibitem[Meyer(2012)]{Meyer2011}
Meyer,~H.-D. {Studying molecular quantum dynamics with the multiconfiguration
  time-dependent Hartree method}. \emph{Wiley Interdiscip. Rev.: Comput. Mol.
  Sci.} \textbf{2012}, \emph{2}, 351--374\relax
\mciteBstWouldAddEndPuncttrue
\mciteSetBstMidEndSepPunct{\mcitedefaultmidpunct}
{\mcitedefaultendpunct}{\mcitedefaultseppunct}\relax
\EndOfBibitem
\bibitem[Brommer \latin{et~al.}(2015)Brommer, Kiselev, Schopf, Beck, Roth, and
  Trebin]{Brommer2015}
Brommer,~P.; Kiselev,~A.; Schopf,~D.; Beck,~P.; Roth,~J.; Trebin,~H.-R.
  {Classical interaction potentials for diverse materials from \textit{ab
  initio} data: a review of \textit{potfit}}. \emph{Modelling Simul. Mater.
  Sci. Eng.} \textbf{2015}, \emph{23}, 074002\relax
\mciteBstWouldAddEndPuncttrue
\mciteSetBstMidEndSepPunct{\mcitedefaultmidpunct}
{\mcitedefaultendpunct}{\mcitedefaultseppunct}\relax
\EndOfBibitem
\bibitem[Wang and Thoss(2003)Wang, and Thoss]{Wang2003}
Wang,~H.; Thoss,~M. {Multilayer formulation of the multiconfiguration
  time-dependent Hartree theory}. \emph{J. Chem. Phys.} \textbf{2003},
  \emph{119}, 1289--1299\relax
\mciteBstWouldAddEndPuncttrue
\mciteSetBstMidEndSepPunct{\mcitedefaultmidpunct}
{\mcitedefaultendpunct}{\mcitedefaultseppunct}\relax
\EndOfBibitem
\bibitem[Manthe(2008)]{Manthe2008}
Manthe,~U. {A multilayer multiconfigurational time-dependent Hartree approach
  for quantum dynamics on general potential energy surfaces}. \emph{J. Chem.
  Phys.} \textbf{2008}, \emph{128}, 164116\relax
\mciteBstWouldAddEndPuncttrue
\mciteSetBstMidEndSepPunct{\mcitedefaultmidpunct}
{\mcitedefaultendpunct}{\mcitedefaultseppunct}\relax
\EndOfBibitem
\bibitem[Maiolo \latin{et~al.}(2021)Maiolo, Worth, and
  Burghardt]{Burghardt2021}
Maiolo,~F.~D.; Worth,~G.~A.; Burghardt,~I. {Multi-layer Gaussian-based
  multi-configuration time-dependent Hartree (ML-GMCTDH) simulations of
  ultrafast charge separation in a donor–acceptor complex}. \emph{J. Chem.
  Phys.} \textbf{2021}, \emph{154}, 144106\relax
\mciteBstWouldAddEndPuncttrue
\mciteSetBstMidEndSepPunct{\mcitedefaultmidpunct}
{\mcitedefaultendpunct}{\mcitedefaultseppunct}\relax
\EndOfBibitem
\bibitem[Kurashige(2018)]{Kurashige2018}
Kurashige,~Y. {Matrix product state formulation of the multiconfiguration
  time-dependent Hartree theory}. \emph{J. Chem. Phys.} \textbf{2018},
  \emph{149}, 194114\relax
\mciteBstWouldAddEndPuncttrue
\mciteSetBstMidEndSepPunct{\mcitedefaultmidpunct}
{\mcitedefaultendpunct}{\mcitedefaultseppunct}\relax
\EndOfBibitem
\bibitem[Li \latin{et~al.}(2020)Li, Ren, and Shuai]{Shuai2020}
Li,~W.; Ren,~J.; Shuai,~Z. {Numerical assessment for accuracy and GPU
  acceleration of TD-DMRG time evolution schemes}. \emph{J. Chem. Phys.}
  \textbf{2020}, \emph{152}, 024127\relax
\mciteBstWouldAddEndPuncttrue
\mciteSetBstMidEndSepPunct{\mcitedefaultmidpunct}
{\mcitedefaultendpunct}{\mcitedefaultseppunct}\relax
\EndOfBibitem
\bibitem[Shepard(1968)]{Shepard1968}
Shepard,~D. {A two-dimensional interpolation function for irregularly-spaced
  data}. {Proceedings of the 1968 23rd ACM National Conference}. 1968; pp
  517--524\relax
\mciteBstWouldAddEndPuncttrue
\mciteSetBstMidEndSepPunct{\mcitedefaultmidpunct}
{\mcitedefaultendpunct}{\mcitedefaultseppunct}\relax
\EndOfBibitem
\bibitem[Haag \latin{et~al.}(2011)Haag, Marti, and Reiher]{haag2011}
Haag,~M.~P.; Marti,~K.~H.; Reiher,~M. {Generation of Potential Energy Surfaces
  in High Dimensions and Their Haptic Exploration}. \emph{ChemPhysChem}
  \textbf{2011}, \emph{12}, 3204--3213\relax
\mciteBstWouldAddEndPuncttrue
\mciteSetBstMidEndSepPunct{\mcitedefaultmidpunct}
{\mcitedefaultendpunct}{\mcitedefaultseppunct}\relax
\EndOfBibitem
\bibitem[Behler and Parrinello(2007)Behler, and Parrinello]{Parrinello2007}
Behler,~J.; Parrinello,~M. {Generalized neural-network representation of
  high-dimensional potential-energy surfaces}. \emph{Phys. Rev. Lett.}
  \textbf{2007}, \emph{98}, 146401\relax
\mciteBstWouldAddEndPuncttrue
\mciteSetBstMidEndSepPunct{\mcitedefaultmidpunct}
{\mcitedefaultendpunct}{\mcitedefaultseppunct}\relax
\EndOfBibitem
\bibitem[Braams and Bowman(2009)Braams, and Bowman]{Bowman2009}
Braams,~B.~J.; Bowman,~J.~M. {Permutationally invariant potential energy
  surfaces in high dimensionality}. \emph{Int. Rev. Phys. Chem.} \textbf{2009},
  \emph{28}, 577--606\relax
\mciteBstWouldAddEndPuncttrue
\mciteSetBstMidEndSepPunct{\mcitedefaultmidpunct}
{\mcitedefaultendpunct}{\mcitedefaultseppunct}\relax
\EndOfBibitem
\bibitem[Dral \latin{et~al.}(2020)Dral, Owens, Dral, and
  Cs{\'{a}}nyi]{Dral2020}
Dral,~P.~O.; Owens,~A.; Dral,~A.; Cs{\'{a}}nyi,~G. {Hierarchical machine
  learning of potential energy surfaces}. \emph{J. Chem. Phys.} \textbf{2020},
  \emph{152}, 204110\relax
\mciteBstWouldAddEndPuncttrue
\mciteSetBstMidEndSepPunct{\mcitedefaultmidpunct}
{\mcitedefaultendpunct}{\mcitedefaultseppunct}\relax
\EndOfBibitem
\bibitem[Li \latin{et~al.}(2019)Li, Song, and Behler]{Behler2019}
Li,~J.; Song,~K.; Behler,~J. {A critical comparison of neural network
  potentials for molecular reaction dynamics with exact permutation symmetry}.
  \emph{Phys. Chem. Chem. Phys.} \textbf{2019}, \emph{21}, 9672--9682\relax
\mciteBstWouldAddEndPuncttrue
\mciteSetBstMidEndSepPunct{\mcitedefaultmidpunct}
{\mcitedefaultendpunct}{\mcitedefaultseppunct}\relax
\EndOfBibitem
\bibitem[Conte \latin{et~al.}(2020)Conte, Qu, Houston, and Bowman]{Conte2020}
Conte,~R.; Qu,~C.; Houston,~P.~L.; Bowman,~J.~M. {Efficient generation of
  permutationally invariant potential energy surfaces for large molecules}.
  \emph{J. Chem. Theory Comput.} \textbf{2020}, \emph{16}, 3264--3272\relax
\mciteBstWouldAddEndPuncttrue
\mciteSetBstMidEndSepPunct{\mcitedefaultmidpunct}
{\mcitedefaultendpunct}{\mcitedefaultseppunct}\relax
\EndOfBibitem
\bibitem[Behler(2015)]{Behler2015}
Behler,~J. {Constructing high-dimensional neural network potentials: A tutorial
  review}. \emph{Int. J. Quantum Chem.} \textbf{2015}, \emph{115},
  1032--1050\relax
\mciteBstWouldAddEndPuncttrue
\mciteSetBstMidEndSepPunct{\mcitedefaultmidpunct}
{\mcitedefaultendpunct}{\mcitedefaultseppunct}\relax
\EndOfBibitem
\bibitem[Amabilino \latin{et~al.}(2019)Amabilino, Bratholm, Bennie, Vaucher,
  Reiher, and Glowacki]{Amabilino2019}
Amabilino,~S.; Bratholm,~L.~A.; Bennie,~S.~J.; Vaucher,~A.~C.; Reiher,~M.;
  Glowacki,~D.~R. {Training Neural Nets To Learn Reactive Potential Energy
  Surfaces Using Interactive Quantum Chemistry in Virtual Reality}. \emph{J.
  Phys. Chem. A} \textbf{2019}, \emph{123}, 4486--4499\relax
\mciteBstWouldAddEndPuncttrue
\mciteSetBstMidEndSepPunct{\mcitedefaultmidpunct}
{\mcitedefaultendpunct}{\mcitedefaultseppunct}\relax
\EndOfBibitem
\bibitem[Richings \latin{et~al.}(2019)Richings, Robertson, and
  Habershon]{Habershon2019}
Richings,~G.~W.; Robertson,~C.; Habershon,~S. {Improved on-the-Fly MCTDH
  Simulations with Many-Body-Potential Tensor Decomposition and Projection
  Diabatization}. \emph{J. Chem. Theory Comput.} \textbf{2019}, \emph{15},
  857--870\relax
\mciteBstWouldAddEndPuncttrue
\mciteSetBstMidEndSepPunct{\mcitedefaultmidpunct}
{\mcitedefaultendpunct}{\mcitedefaultseppunct}\relax
\EndOfBibitem
\bibitem[Manzhos and Carrington(2006)Manzhos, and Carrington]{Manzhos2006}
Manzhos,~S.; Carrington,~T. {Using neural networks to represent potential
  surfaces as sums of products}. \emph{J. Chem. Phys.} \textbf{2006},
  \emph{125}, 194105\relax
\mciteBstWouldAddEndPuncttrue
\mciteSetBstMidEndSepPunct{\mcitedefaultmidpunct}
{\mcitedefaultendpunct}{\mcitedefaultseppunct}\relax
\EndOfBibitem
\bibitem[Koch and Zhang(2014)Koch, and Zhang]{Koch2014}
Koch,~W.; Zhang,~D.~H. {Communication: Separable potential energy surfaces from
  multiplicative artificial neural networks}. \emph{J. Chem. Phys.}
  \textbf{2014}, \emph{141}, 021101\relax
\mciteBstWouldAddEndPuncttrue
\mciteSetBstMidEndSepPunct{\mcitedefaultmidpunct}
{\mcitedefaultendpunct}{\mcitedefaultseppunct}\relax
\EndOfBibitem
\bibitem[Westermayr \latin{et~al.}(2019)Westermayr, Gastegger, Menger, Mai,
  Gonz{\'{a}}lez, and Marquetand]{Marquetand2019}
Westermayr,~J.; Gastegger,~M.; Menger,~M. F. S.~J.; Mai,~S.;
  Gonz{\'{a}}lez,~L.; Marquetand,~P. {Machine learning enables long time scale
  molecular photodynamics simulations}. \emph{Chem. Sci.} \textbf{2019},
  \emph{10}, 8100--8107\relax
\mciteBstWouldAddEndPuncttrue
\mciteSetBstMidEndSepPunct{\mcitedefaultmidpunct}
{\mcitedefaultendpunct}{\mcitedefaultseppunct}\relax
\EndOfBibitem
\bibitem[Dral \latin{et~al.}(2018)Dral, Barbatti, and Thiel]{Thiel2018}
Dral,~P.~O.; Barbatti,~M.; Thiel,~W. {Nonadiabatic Excited-State Dynamics with
  Machine Learning}. \emph{J. Phys. Chem. Lett.} \textbf{2018}, \emph{9},
  5660--5663\relax
\mciteBstWouldAddEndPuncttrue
\mciteSetBstMidEndSepPunct{\mcitedefaultmidpunct}
{\mcitedefaultendpunct}{\mcitedefaultseppunct}\relax
\EndOfBibitem
\bibitem[Baiardi and Reiher(2020)Baiardi, and Reiher]{Baiardi2020}
Baiardi,~A.; Reiher,~M. {The density matrix renormalization group in chemistry
  and molecular physics: Recent developments and new challenges}. \emph{J.
  Chem. Phys.} \textbf{2020}, \emph{152}, 040903\relax
\mciteBstWouldAddEndPuncttrue
\mciteSetBstMidEndSepPunct{\mcitedefaultmidpunct}
{\mcitedefaultendpunct}{\mcitedefaultseppunct}\relax
\EndOfBibitem
\bibitem[Polyak \latin{et~al.}(2019)Polyak, Richings, Habershon, and
  Knowles]{Habershon2019a}
Polyak,~I.; Richings,~G.~W.; Habershon,~S.; Knowles,~P.~J. {Direct quantum
  dynamics using variational Gaussian wavepackets and Gaussian process
  regression}. \emph{J. Chem. Phys.} \textbf{2019}, \emph{150}, 041101\relax
\mciteBstWouldAddEndPuncttrue
\mciteSetBstMidEndSepPunct{\mcitedefaultmidpunct}
{\mcitedefaultendpunct}{\mcitedefaultseppunct}\relax
\EndOfBibitem
\bibitem[Mart{\'i}nez-N{\'u}{\~n}ez(2015)]{Martinez2015}
Mart{\'i}nez-N{\'u}{\~n}ez,~E. {An Automated Method to Find Transition States
  Using Chemical Dynamics Simulations}. \emph{J. Comput. Chem.} \textbf{2015},
  \emph{36}, 222--234\relax
\mciteBstWouldAddEndPuncttrue
\mciteSetBstMidEndSepPunct{\mcitedefaultmidpunct}
{\mcitedefaultendpunct}{\mcitedefaultseppunct}\relax
\EndOfBibitem
\bibitem[D{\"o}ntgen \latin{et~al.}(2015)D{\"o}ntgen, {Przybylski-Freund},
  Kr{\"o}ger, Kopp, Ismail, and Leonhard]{Dontgen2015}
D{\"o}ntgen,~M.; {Przybylski-Freund},~M.-D.; Kr{\"o}ger,~L.~C.; Kopp,~W.~A.;
  Ismail,~A.~E.; Leonhard,~K. {Automated Discovery of Reaction Pathways, Rate
  Constants, and Transition States Using Reactive Molecular Dynamics
  Simulations}. \emph{J. Chem. Theory Comput.} \textbf{2015}, \emph{11},
  2517--2524\relax
\mciteBstWouldAddEndPuncttrue
\mciteSetBstMidEndSepPunct{\mcitedefaultmidpunct}
{\mcitedefaultendpunct}{\mcitedefaultseppunct}\relax
\EndOfBibitem
\bibitem[Wu \latin{et~al.}(2019)Wu, Sun, Wu, and Deetz]{Wu2019}
Wu,~Y.; Sun,~H.; Wu,~L.; Deetz,~J.~D. {Extracting the Mechanisms and Kinetic
  Models of Complex Reactions from Atomistic Simulation Data}. \emph{J. Comput.
  Chem.} \textbf{2019}, \emph{40}, 1586--1592\relax
\mciteBstWouldAddEndPuncttrue
\mciteSetBstMidEndSepPunct{\mcitedefaultmidpunct}
{\mcitedefaultendpunct}{\mcitedefaultseppunct}\relax
\EndOfBibitem
\bibitem[Liu \latin{et~al.}(2014)Liu, Li, Guo, Zheng, Han, Yuan, Nie, and
  Liu]{Liu2014}
Liu,~J.; Li,~X.; Guo,~L.; Zheng,~M.; Han,~J.; Yuan,~X.; Nie,~F.; Liu,~X.
  {Reaction analysis and visualization of ReaxFF molecular dynamics
  simulations}. \emph{J. Mol. Graph. Modell.} \textbf{2014}, \emph{53},
  13--22\relax
\mciteBstWouldAddEndPuncttrue
\mciteSetBstMidEndSepPunct{\mcitedefaultmidpunct}
{\mcitedefaultendpunct}{\mcitedefaultseppunct}\relax
\EndOfBibitem
\bibitem[Hutchings \latin{et~al.}(2020)Hutchings, Liu, Qiu, Song, and
  Wang]{Hutchings2020}
Hutchings,~M.; Liu,~J.; Qiu,~Y.; Song,~C.; Wang,~L.-P. {Bond-Order Time Series
  Analysis for Detecting Reaction Events in \textit{Ab Initio} Molecular
  Dynamics Simulations}. \emph{J. Chem. Theory Comput.} \textbf{2020},
  \emph{16}, 1606--1617\relax
\mciteBstWouldAddEndPuncttrue
\mciteSetBstMidEndSepPunct{\mcitedefaultmidpunct}
{\mcitedefaultendpunct}{\mcitedefaultseppunct}\relax
\EndOfBibitem
\bibitem[Hirai and Jinnouchi(2021)Hirai, and Jinnouchi]{Hirai2021}
Hirai,~H.; Jinnouchi,~R. {Discovering chemical reaction pathways using
  accelerated molecular dynamics simulations and network analysis tools ---
  Application to oxidation induced decomposition of ethylene carbonate}.
  \emph{Chem. Phys. Lett.} \textbf{2021}, \emph{770}, 138439\relax
\mciteBstWouldAddEndPuncttrue
\mciteSetBstMidEndSepPunct{\mcitedefaultmidpunct}
{\mcitedefaultendpunct}{\mcitedefaultseppunct}\relax
\EndOfBibitem
\bibitem[Miller \latin{et~al.}(1983)Miller, Schwartz, and Tromp]{Miller1983}
Miller,~W.~H.; Schwartz,~S.~D.; Tromp,~J.~W. {Quantum mechanical rate constants
  for bimolecular reactions}. \emph{J. Chem. Phys.} \textbf{1983}, \emph{79},
  4889--4898\relax
\mciteBstWouldAddEndPuncttrue
\mciteSetBstMidEndSepPunct{\mcitedefaultmidpunct}
{\mcitedefaultendpunct}{\mcitedefaultseppunct}\relax
\EndOfBibitem
\bibitem[Richardson(2018)]{Richardson2018}
Richardson,~J.~O. {Ring-polymer instanton theory}. \emph{Int. Rev. Phys. Chem.}
  \textbf{2018}, \emph{37}, 171--216\relax
\mciteBstWouldAddEndPuncttrue
\mciteSetBstMidEndSepPunct{\mcitedefaultmidpunct}
{\mcitedefaultendpunct}{\mcitedefaultseppunct}\relax
\EndOfBibitem
\end{mcitethebibliography}
\end{document}